\documentclass[prd,nofootinbib,preprint,superscriptaddress,twocolumn,10pt]{revtex4}
\pdfoutput=1
\usepackage{amsmath}
\usepackage{float}
\usepackage{amsfonts}
\usepackage{amssymb}
\usepackage{mathrsfs}
\usepackage{cancel}
\usepackage{accents}
\usepackage{mciteplus,slashed}
\usepackage{amssymb,cancel,amsmath,relsize}
\usepackage{mathrsfs} 
\usepackage{dcolumn}
\usepackage{bm}
\usepackage[caption=false]{subfig} 
\usepackage{appendix}
\usepackage{physics}
\usepackage{feynmp-auto}
\usepackage[T1]{fontenc}	
\usepackage{csvsimple}
\usepackage{hyperref}
\usepackage[capitalise]{cleveref}
\usepackage{booktabs}
\usepackage{graphicx}
\usepackage{mathrsfs}
\usepackage[utf8]{inputenc}
\usepackage[T1]{fontenc}
\usepackage[dvipsnames]{xcolor}

\hypersetup{
    colorlinks,
    linkcolor={red!50!black},
    citecolor={blue!50!black},
    urlcolor={blue!80!black}
}
\usepackage[normalem]{ulem}
\usepackage{cleveref}

\renewcommand{\section}[1]{{\noindent \bf{#1.}---}}

\newcommand{\bmt}{\begin{pmatrix}}
\newcommand{\emt}{\end{pmatrix}}
\newcommand{\ba}{\begin{array}{c}}
\newcommand{\ea}{\end{array}}
\newcommand{\be}{\begin{equation}}
\newcommand{\ee}{\end{equation}}
\newcommand{\bea}{\begin{eqnarray}}
\newcommand{\eea}{\end{eqnarray}}

\newcommand{\bi}{\begin{itemize}}
\newcommand{\ei}{\end{itemize}}

\newcommand{\baz}{\begin{array}{cc}}
\newcommand{\besub}{\begin{subequations}}
\newcommand{\eesub}{\end{subequations}}

\newcommand{\mathsym}[1]{{}}

\newcommand{\bt}{\begin{tabular}}
\newcommand{\et}{\end{tabular}}

\newcommand{\benu}{\begin{enumerate}}
\newcommand{\eenu}{\end{enumerate}}

\def\q2 {q^2}

\def\bt{\begin{table}}
\def\et{\end{table}}

\newcommand{\bav}{\begin{array}{cccc}}

\graphicspath{{Figures/}}

\begin{document}

\title{Imprint of domain wall annihilation on induced gravitational waves }

\author{Rishav Roshan}
\affiliation{School of Physics and Astronomy, University of Southampton, Southampton SO17 1BJ, United Kingdom}

\begin{abstract}
Domain wall annihilation can leave a distinctive imprint on the induced gravitational wave spectrum.  During annihilation, most of the domain wall energy transforms into the scalar field responsible for the initial 
$\mathbb{Z}_2$ symmetry breaking that created the walls, along with any coupled species. If the produced scalar is sufficiently long-lived, its delayed decay drives an early matter-dominated phase following domain wall annihilation, significantly amplifying induced gravitational waves from primordial perturbations. The subsequent transition to radiation domination dilutes the domain wall contribution through entropy injection while preserving the enhanced induced signal.  This creates a gravitational wave spectrum with two distinct peaks detectable across complementary frequency bands. We explore the observable parameter space and demonstrate how multi-band detection can probe early universe symmetry breaking.
\end{abstract}

\maketitle
\tableofcontents
\section{Introduction}

The recent discovery of stochastic gravitational wave background (SGWB), as reported by the North American Nanohertz Observatory for Gravitational Waves (NANOGrav)~\cite{NANOGrav:2023gor, NANOGrav:2023hvm} has now  motivated us to explore of cosmological events from the early Universe that could generate such a background and potentially reveal physics beyond the Standard Model (SM).

The origin of SGWBs can be associated to a variety of early cosmological events, such as inflation \cite{Grishchuk:1974ny,Starobinsky:1979ty,Guzzetti:2016mkm} (primordial tensor \cite{Giovannini:1998bp,Giovannini:1999bh,Riazuelo:2000fc,Seto:2003kc,Boyle:2007zx,Gouttenoire:2021jhk,Haque:2021dha,Chakraborty:2023ocr,Barman:2022qgt,Barman:2023ktz,Barman:2024ujh,Kitajima:2023kzu} and scalar \cite{Matarrese:1992rp,Matarrese:1993zf,Matarrese:1997ay,Mollerach:2003nq,Ananda:2006af,Baumann:2007zm,Domenech:2021ztg,Roshan:2024qnv} fluctuations), first-order phase transitions~\cite{Winicour1973, Hogan:1986qda, Athron:2023xlk, Caprini:2010xv, NANOGrav:2021flc, Xue:2021gyq, DiBari:2021dri, Madge:2023cak,Mandal:2026wbz}, collapse of topological defects like cosmic strings~\cite{Siemens:2006yp, Ellis:2020ena, King:2020hyd, Buchmuller:2020lbh, Blasi:2020mfx, Bian:2022tju, Fu:2023nrn, Ghoshal:2025iil}, or domain walls (DWs)~\cite{Ferreira:2022zzo, An:2023idh, Dunsky:2021tih, Sakharov:2021dim,Bai:2023cqj}, primordial black holes (PBHs) dynamics \cite{Domenech:2024kmh, Flores:2024eyy, del-Corral:2025fca, Gross:2024wkl,Borah:2022vsu,Barman:2022pdo,Borah:2023iqo,Barman:2024slw,Borah:2024lml} etc. Existing literature mostly treats these sources in isolation, analysing their individual contributions to the GW spectrum. However, cosmological histories involving multiple epochs can produce correlated signals where one phenomenon affects or enhances signatures from another. In this context, we examine a previously unexplored connection: the imprint left by DW's annihilation on induced gravitational waves (IGWs) sourced by second-order scalar \cite{Matarrese:1992rp,Matarrese:1993zf,Matarrese:1997ay,Mollerach:2003nq,Ananda:2006af,Baumann:2007zm,Domenech:2021ztg,Roshan:2024qnv} perturbations.

Phase transitions that involve spontaneous breaking of a discrete symmetry give rise to a sheet like topological defect known as DWs \cite{Vilenkin:1984ib,Kolb:1990vq,Preskill:1992ck,1994csot.book.....V}. The study of DWs has traditionally focused on second-order phase transitions (SOPTs), where they appear after a smooth phase transition, an idea originally introduced by T. D. Lee~\cite{Lee:1974jb} \footnote{For first-order phase transitions involving $\mathbb{Z}_2$ breaking that result in a DW formation see \cite{Wei:2022poh,Borboruah:2022eex,Fornal:2024avx,Roshan:2026xpf}}. Once formed, their energy density dilutes slower ($\rho_{\rm DW}\propto t^{-1}$) in comparison to that of radiation and matter, causing them to rapidly dominate the energy budget of the Universe. This famous DW problem can be resolved by destabilizing the walls through explicit breaking of the discrete symmetry \cite{Gelmini:1988sf} and introducing biased conditions at the time of formation\footnote{However, this bias can also be generated through radiative corrections~\cite{Zhang:2023nrs, Zeng:2025zjp, Borah:2025bfa,Borah:2026kfo}.}. This allows the DW network to annihilate prior to dominating the Universe's energy budget, efficiently producing GWs in the process~\cite{Saikawa:2017hiv,Roshan:2024qnv,King:2023ayw,King:2023ztb,Bhattacharya:2023kws,Borah:2024kfn, Gouttenoire:2025ofv,Kitajima:2023cek,Vanvlasselaer:2026fay}. Moreover, the DW collapse also results in the particle production. A significant amount of the DW energy is transferred to the scalar field (whose $\mathbb{Z}_2$ symmetry breaking produced the DWs) and any other particles coupled to it \cite{Widrow:1989vj}. Interestingly, if the produced scalar is sufficiently long-lived, its delayed decay drives an early matter-dominated (MD) phase following DW annihilation. In this work, we explore the cosmological consequences of this scenario in context of GWs signatures. 

During the inflationary phase, scalar curvature perturbations can be developed in the early Universe \cite{Matarrese:1992rp,Matarrese:1993zf,Matarrese:1997ay,Mollerach:2003nq,Ananda:2006af,Baumann:2007zm,Domenech:2021ztg,Roshan:2024qnv}. While at first order, these scalar modes cannot directly produce the transverse-traceless tensor modes that constitute GWs, extending perturbation theory to second order reveals that products of first-order scalar perturbations can source tensor modes. As modes enter the horizon during radiation domination, scalar perturbations interact at second-order to produce GWs. The resulting IGW spectrum carries information about small-scale scalar perturbations and provides an independent window into primordial Universe. Interestingly, if the Universe proceeds through an era of early MD following the initial radiation domination after reheating, the production of such IGWs gets intensified \cite{Assadullahi:2009nf,Baumann:2007zm,Jedamzik:2010dq,Alabidi:2013lya,Kohri:2018awv,Inomata:2019ivs,Inomata:2019zqy,Papanikolaou:2020qtd,Domenech:2020ssp,Dalianis:2020gup,Domenech:2021ztg,Das:2021wad,Eggemeier:2022gyo,Fernandez:2023ddy,Chianese:2025mll} but the accurate estimation of such spectrum are challenging \cite{Fernandez:2023ddy, Chianese:2025mll}. The degree of enhancement depends on whether perturbations evolve linearly or non-linearly. In the linear regime \cite{Kohri:2018awv,Inomata:2019ivs,Inomata:2019zqy}, where perturbations remain small and analytical or semi-analytical methods reliably estimate IGW production, one expects moderate enhancement of the IGW spectrum. However, significantly larger amplification occurs in the non-linear regime \cite{Jedamzik:2010dq,Fernandez:2023ddy}, where perturbations grow sufficiently to form non-linear structures. As a result, non-linear regime requires numerical simulations for accurate predictions \cite{Eggemeier:2022gyo,Fernandez:2023ddy, Dalianis:2024kjr}. 

In this work, we focus on the non-linear regime and explore a previously unexamined cosmological scenario linking DW dynamics to the enhancement of IGWs. Specifically, DW annihilation generates both a direct SGWB from topological defect collapse and a population of scalar particles whose long lifetime enables them to dominate the cosmic energy density before eventually decaying. This intermediate MD era plays a critical role in reshaping the GW landscape \footnote{The presence of early MD era can also modifies the GW from different primordial sources as shown in \cite{Ghoshal:2026ros}. Moreover, presence of such era can also leave observable imprints on other cosmological phenomena as discussed in \cite{Roshan:2025mwl,Konar:2025gvh,Datta:2025vyu} }: scalar curvature perturbations entering the horizon during MD source IGWs with dramatically enhanced amplitude compared to radiation-dominated production. When the scalar field finally decays, the transition back to radiation domination introduces differential effects on the two GW populations, entropy injection from decay dilutes the DW's GW signals, while preserving the enhanced induced signal. We demonstrate that this interplay results in a distinctive GW spectrum: a higher-frequency peak directly reflecting DW annihilation dynamics, and a lower-frequency peak carrying the imprint of enhanced IGWs, together providing complementary observational windows detectable by multi-band GW experiments. For our analysis of IGWs during the early MD era, we adopt the framework of \cite{Fernandez:2023ddy}, which employs hybrid N-body and lattice simulations to track the formation and decay of early structures and compute the resulting IGW spectrum.

The paper is organized as follows. In section \ref{DWPH} we discuss the DW phenomenology. In section \ref{EMD} we discuss how the produced scalar from DW annihilation induces an era early MD. In section \ref{GW_analysis} we do the GW analysis, following which we summarize our results in section \ref{sec:summary}. Finally, we conclude in Section~\ref{conclusion}.


\section{Domain wall phenomenology}
\label{DWPH}

\subsection{Toy model}
\label{model}
We consider a toy model with a minimal extension of the SM by adding a real scalar field $S$ that transforms under a $\mathbb{Z}_2$ symmetry as $ S\to -S$. The scalar breaks the  $\mathbb{Z}_2$ symmetry spontaneously when it obtains a non-zero vacuum expectation value (VEV), $v_S$. The most general renormalizable scalar potential involving the SM Higgs doublet $H$ and the $\mathbb{Z}_2$-odd real scalar singlet $S$ at tree level can then be expressed as:
\begin{eqnarray}
    V&=& \frac{\mu ^2}{2}  H^\dagger H+ \frac{\lambda_H}{4} (H^\dagger H)^2 + \frac{\lambda _{HS}}{4} H^\dagger H S^2  \nonumber 
    \\ 
    && \hspace{0cm}+\frac{\lambda_S}{4} (S^2-v_S^2)^2  \; .
    \label{eq:lagrangian_tree}
\end{eqnarray}
One should note that the scalar potential should be bounded from below to make the electroweak vacuum stable, which leads to the following constraints~\cite{Kannike:2012pe,Chakrabortty:2013mha}:
\begin{eqnarray}
    \lambda_H,\lambda_S\geq 0 \; ,\qquad \lambda_{HS}+2\sqrt{\lambda_H\lambda_S}\geq 0 \;.
\end{eqnarray}

\noindent A small quartic coupling $\lambda_{HS}$ between $S$ and the SM Higgs, together with the large hierarchy $v_{S}\gg v_h$, ensures that the mixing angle $\theta_{HS} \sim \lambda_{HS} v_h v_{S}/(m_{S}^2 - m^2_h)$ remains sufficiently small after Higgs field acquires its VEV $v_h$. This also prevent large quadratic correction to the Higgs mass, avoids observable deviations in Higgs phenomenology and allows us to neglect $S$-Higgs mixing in our analysis.

\subsection{A biased domain wall network}

Assuming that the Universe went through a SOPT during the spontaneous breaking of the $\mathbb{Z}_2$ symmetry, $i.e.$ the Universe smoothly evolves from the symmetric phase to the broken phase as the temperature decreases, causality prevents different regions of the Universe from choosing the same vacuum state. At the time of the transition, the Universe is divided into causally disconnected regions with size of order the correlation length $L (T)=1/m_S(T)$, where $m_S(T)$ is thermal mass. In each region, the scalar field independently settles into either $+v_S$ or $-v_S$ with roughly equal probability. When neighboring regions choose different vacua, the scalar field must interpolate between them. This interpolation cannot occur abruptly, instead, the field varies smoothly across a finite thickness region, forming a DW \cite{Zeldovich:1974uw,Kibble:1976sj}. Thus, DWs appear at the boundaries separating regions with different vacuum choices. Once formed, their energy density ($\rho_{\rm DW}$) dilute slowly,

\begin{eqnarray}
    \rho_{\rm DW}=\frac{\sigma}{L}, ~ {\rm{with}}~ L\sim t/\mathcal{A},
    \label{rhoDW}
\end{eqnarray}

\noindent where $\sigma=\sqrt{\frac{8\lambda_S}{9}}v_S^3$ denotes the surface tension of the DW and the area parameter has been calculated to be $\mathcal{A} = 0.8 \pm 0.1$ \cite{Hiramatsu:2013qaa} for the $\mathbb{Z}_2$-symmetric potential. The complex dynamics of DW network can be well described by a scaling law in which the correlation length of the network is set by the cosmic horizon $L\simeq t$ \cite{Press:1989yh}. This slower dilution allow them to rapidly dominate the energy budget of the Universe. In order to avoid such catastrophe, the DWs can be made unstable by explicitly breaking the discrete symmetries \cite{Gelmini:1988sf} at the time of formation. This can be achieved by introducing a term like
 
 \begin{equation}
     \Delta V=\epsilon v_S S \bigg{(}\frac{1}{3}S^2-v_S^2\bigg{)}
 \end{equation}
 
  \noindent in the scalar potential. Here $\epsilon$ is a dimensionless constant. The biasi in the scalar potential can be evaluated by considering $ \Delta V (-v_S)- \Delta V (v_S)$ which gives,
  \begin{equation}
     V_{\rm bias}=\frac{4}{3}\epsilon v_S^4.
     \label{vbias}
 \end{equation}

\noindent Presence of the $V_{\rm bias}$ in Eq.~\eqref{vbias} exerts pressure on the walls. When this pressure exceeds the surface tension pressure $C_{\rm ann}\, \sigma / L$, the DW network annihilates at time
\begin{equation}
    \label{eq:t_ann}
    t_{\rm ann} \simeq  C_{\rm ann} \,\mathcal{A}\,\frac{\sigma}{V_{\rm bias}} \;,
\end{equation}
where $C_{\rm ann}\sim 2$ from lattice simulations \cite{Kawasaki:2014sqa}. Cosmological viability requires annihilation before the DW network dominates, i.e., $t_{\rm ann} \lesssim t_{\rm dom}$, where $t_{\rm dom}$ satisfies
\begin{equation}
\label{eq:t_dom_def}
\mathcal{A}\sigma/t_{\rm dom}+V_{\rm bias} \simeq \rho_{\rm rad}(t_{\rm dom}) \; .
\end{equation}
Approximating $\rho_{\rm rad}\simeq 3M_{\rm pl}^2/(2t)^2$ as in pure radiation domination, we find
\begin{equation}
    \label{eq:t_dom}
     t_{\rm dom } \simeq \frac{\sqrt{3}M_{\rm pl}}{2\sqrt{V_{\rm bias}}} \;,
\end{equation}
neglecting $\mathcal{O}(t_{\rm ann}/t_{\rm dom})$ corrections. The total energy fraction of the DW network plus vacuum bias at annihilation is
\begin{align}
\label{eq:alpha_ann_DW_V}
    \alpha^{\rm DW+V}_{\rm ann} &\simeq \frac{\rho_{\rm DW}+\rho_{\rm V}}{3M_{\rm pl}^2\mathcal{H}^2} \Big|_{t=t_{\rm ann}}\simeq (1+C_d/2)\alpha_{\rm ann}^{\rm DW} \;,
\end{align}
where $\rho_{\rm V}\simeq V_{\rm bias}/2$ is the bulk vacuum energy density. The DW surface energy fraction at annihilation is given by
\begin{align}
    \label{eq:alpha_DW}
     \alpha_{\rm ann}^{\rm DW} \equiv \frac{\rho_{\rm DW}}{3M_{\rm pl}^2\mathcal{H}^2} \Big|_{t=t_{\rm ann}}
     \simeq C_d^{-1}\left( \frac{t_{\rm ann}}{t_{\rm dom }} \right)^2 \;,
\end{align}
using $\mathcal{H}\simeq 1/2t$.
 
Next, we address constraints on $V_{\rm bias}$ arising from DW decay into SM particles. DWs with very small energy bias can exist for long periods, since the annihilation timescale scales inversely with the bias ($t_{\rm ann}\propto V_{\rm bias}^{-1}$). Demanding that DWs collapse before dominating the Universe's energy density establishes a lower bound on $|V_{\rm bias}|$ \cite{Saikawa:2017hiv}.
\begin{equation}
V_{\rm bias}^{1/4}>2.18\times10^{-5} {\rm GeV}~ C_{\rm ann}^{1/4}\mathcal{A}^{1/2}\left(\frac{\sigma^{1/3}}{10^3\rm GeV}\right)^{3/2}.
\label{vbias_LB}
\end{equation}
Satisfying Eq. \eqref{vbias_LB} ensures DWs annihilate before dominating the energy density. However, if DWs decay to SM particles, these decay products can destroy light elements synthesized during BBN. To avoid this, DW annihilation must complete by $t_{\rm ann}\leq 0.01$ s, which imposes a further constraint on $V_{\rm bias}$:
\begin{equation}
V_{\rm bias}^{1/4}>5.07\times10^{-4} {\rm GeV}~ C_{\rm ann}^{1/4}\mathcal{A}^{1/4}\left(\frac{\sigma^{1/3}}{10^3\rm GeV}\right)^{3/4}.
\label{vbias_LB2}
\end{equation}
In addition  to this, percolation theory predicts that an infinite false vacuum cluster forms when its probability exceeds the threshold $p_c = 0.311$ \cite{Saikawa:2017hiv}. This constrains the bias potential to
\begin{equation}
\frac{V_{\rm bias}}{V_0 } < 0.795 \;,
\end{equation}
which ensures DW generation. Here, $V_0$ denotes the height of the potential barrier and for simplicity we consider $V_0=\sigma^{4/3}$ following \cite{Saikawa:2017hiv}.

\begin{figure}[htbp!]
    \centering
    \includegraphics[width=0.8\linewidth]{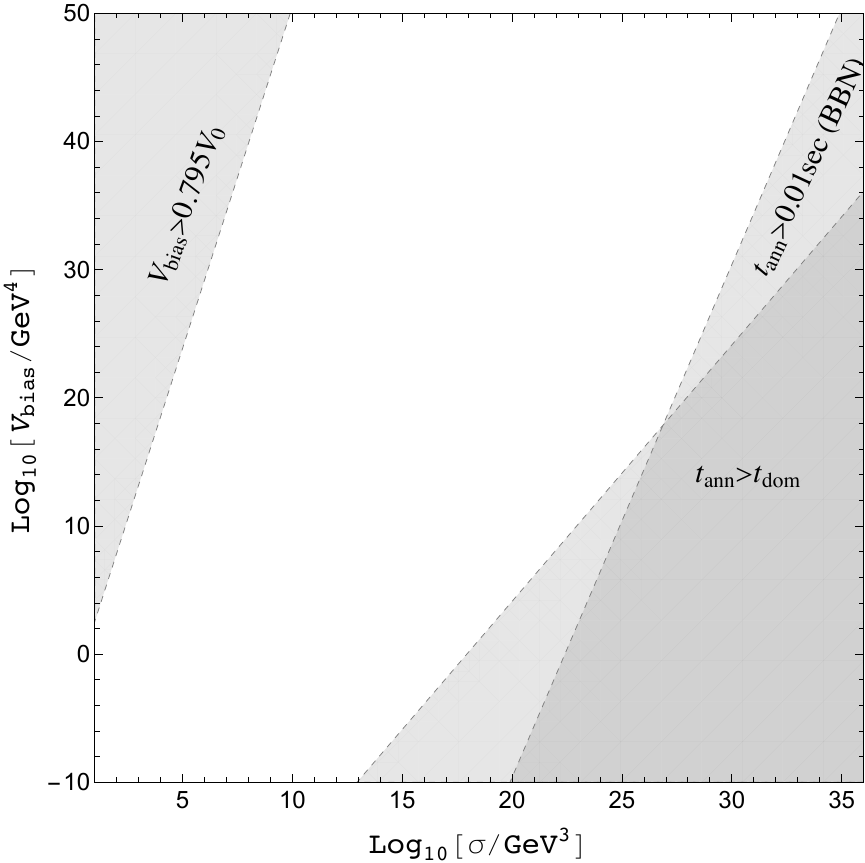}
    \caption{Parameter space in $V_{\rm bias}-\sigma$ plane. Gray shaded regions show the parameter space ruled out by physical constraints. }
    \label{fig:DW_constraints}
\end{figure}

In order to present a model independent analysis, we consider $\sigma$ and $V_{\rm bias}$ as a free parameter and show these constraints in $V_{\rm bias}-\sigma$ plane in Fig. \ref{fig:DW_constraints}. 

\section{Early Matter-Domination Era}
\label{EMD}

 As discussed previously, we consider scenarios where $V_{\rm bias}$ is sufficiently large and DWs collapse instantaneously, well before they can dominate the Universe's energy density. During this collapse, most of the DW energy density is transferred to the scalar field $S$ and any particles coupled to it \cite{Widrow:1989vj}. In this work,  we examine the case where $S$ is long-lived and non-relativistic.

Following the Lagrangian in Eq.~\eqref{eq:lagrangian_tree}, $S$ can dominantly decay to SM Higgs boson if kinematically allowed ($m_S\geq 2 m_h$). The decay width of the singlet scalar is then given by:
\begin{align}
    \Gamma_{S} &\simeq \frac{\lambda_{HS}^2v_S^2}{32\pi m_S} \; ,
\end{align}
where $m_S^2=2\lambda_Sv_S^2$. A long-lived $S$ is naturally realized in the current setup through a small Higgs portal coupling $\lambda_{HS}$ (see section \ref{model}).

The energy density of a long-lived, non-relativistic scalar $S$  can be expressed in terms of its comoving number density $Y$ as,
\begin{eqnarray}
    \rho_S(T)= m_S Y s(T),
    \label{rhoS}
\end{eqnarray}
where $s$ denotes the entropy density given by $s(T)=\frac{2\pi^2}{45} g_{s}T^3$ with $g_s$ denoting the effective entropic degrees of freedom and $T$ denoting the temperature of the Universe. On the other hand, the energy density of the radiation at any temperature is defined as
\begin{eqnarray}
    \rho_{R}(T)=\frac{\pi^2}{30} g_{*}T^4,
    \label{rhoR}
\end{eqnarray}
where $g_*$ denotes the effective number of relativistic degrees of freedom. Following Eq. \eqref{rhoS} and \eqref{rhoR} one obtains,

\begin{eqnarray}
    \frac{\rho_S(T)}{\rho_R(T)}=\frac{4}{3}\frac{m_S Y}{T},
    \label{ratio1}
\end{eqnarray}
where we have consider $g_*=g_s$.

Next, the temperature, $T_{\rm S-dom}$ at which $S$ starts to dominate can be calculated by equating $\rho_S(T_{\rm S-dom}) = \rho_R(T_{\rm S-dom})$. Therefore, setting $\frac{\rho_S(T_{\rm S-dom})}{\rho_R(T_{\rm S-dom})}=1$, gives,

\begin{eqnarray}
    T_{\rm S-dom}=\frac{4}{3}m_S Y.
\end{eqnarray}
Further, substituting $T_{\rm S-dom}$ back in Eq. \eqref{ratio1} we find a relation

\begin{eqnarray}
     \frac{\rho_S(T)}{\rho_R(T)}=\frac{T_{\rm S-dom}}{T},
     \label{ratio}
\end{eqnarray}

\noindent where $T_{\textrm{$S$-dom}}$ can be obtained from the relation, 
\begin{equation}
\label{eq:T_S_dom}
   T_{\textrm{$S$-dom}}\simeq \alpha_{\rm ann}^{\rm DW+V}T_{\rm ann}\;  .
\end{equation}

\subsection{$S$ decay  }
We now move on to calculate the decay temperature of $S$, $T_{\rm dec}$. Once the Universe is dominated by the field $S$, the expansion rate $\mathcal{H}$ can be expressed as

\begin{eqnarray}
    \mathcal{H}(T)=\sqrt{\frac{\rho_S(T)+\rho_R(T)}{3M_{p}}}
\end{eqnarray}
where $M_p$ denotes the reduced Planck mass. Using Eq. \eqref{rhoR} and \eqref{ratio}, the Hubble rate can then be expressed as,

\begin{eqnarray}
     \mathcal{H}(T)&=&0.33\sqrt{g_*}\frac{T^2}{M_P}\sqrt{1+\frac{T_{\rm S-dom}}{T}}.
\end{eqnarray}
Assuming $S$ decays instantaneously, one can simply equate $\mathcal{H}(T_{\rm dec})\simeq\Gamma_S$. This gives,
\begin{eqnarray}
 0.33\sqrt{g_*}\frac{T_{\rm dec}^2}{M_P}\sqrt{1+\frac{T_{\rm S-dom}}{T_{\rm dec}}}\simeq\Gamma_S.   
\end{eqnarray}

\noindent Since $S$ starts dominating the Universe much before its decay,   $i.e~T_{\rm S-dom}>>T_{\rm dec}$, the decay temperature can be calculated from

\begin{eqnarray}
    T_{\rm dec}\simeq\bigg{(}\frac{1}{0.33\sqrt{g_*}}\frac{\Gamma_S^2M_P^2}{T_{\rm S-dom}}\bigg)^{1/3}.
\end{eqnarray}
\subsection{Entropy injection from $S$ decay}
Let $\mathcal{S}$ denotes the entropy per comoving volume,
\begin{equation}
    \mathcal{S}(T)=a^3\frac{\rho_R(T)}{T}
\end{equation}
where $a$ denotes the scale factor. Since both entropy
before and after decay is calculated at the time of decay,
we write,
\begin{eqnarray}
    \mathcal{S}_{\rm before}=a_{\rm dec}^3\frac{\rho_{R}^{\rm before}(T_{\rm dec})}{T_{\rm dec}},
\end{eqnarray}
where $\rho_R^{\rm before}$ denotes the radiation energy density (sub-dominant in comparison to $\rho_S$) just before decay. Next we write entropy of the Universe after the decay,
\begin{eqnarray}
    \mathcal{S}_{\rm after}=\mathcal{S}_{\rm before}+\Delta \mathcal{S},
    \label{totalS}
\end{eqnarray}
where $\Delta \mathcal{S}=a_{\rm dec}^3\frac{\rho_R^{\rm after}(T_{\rm dec})}{T_{\rm dec}}$ is the entropy injected to the bath from the $S$ decay. We also assume the entropy injection to be instantaneous. Following Eq. \eqref{totalS} we write,
\begin{eqnarray}
    \mathcal{D}=\frac{\mathcal{S}_{\rm after}}{\mathcal{S}_{\rm before}}=1+\frac{\Delta \mathcal{S}}{\mathcal{S}_{\rm before}},
    \label{D1}
\end{eqnarray}
where $\mathcal{D}$ denotes the dilution factor. As the entropy injection is assumed to be instantaneous, the total energy density of $S$ will be transferred to the bath $i.e~\rho_S=\rho_R^{\rm after}$. Substituting this back in Eq. \eqref{totalS}, we get,
\begin{eqnarray}
    \mathcal{S}_{\rm after}&=&a_{\rm dec}^3\frac{\rho_{R}^{\rm before}(T_{\rm dec})}{T_{\rm dec}}+a_{\rm dec}^3\frac{\rho_R^{\rm after}(T_{\rm dec})}{T_{\rm dec}},\\
    \frac{\mathcal{S}_{\rm after}}{\mathcal{S}_{\rm before}}&=&1+\frac{\rho_{R}^{\rm after}(T_{\rm dec})}{\rho_R^{\rm before}(T_{\rm dec})}
\end{eqnarray}
or,
\begin{eqnarray}
    \mathcal{D}&=&1+\frac{\rho_{s}(T_{\rm dec})}{\rho_R^{\rm before}(T_{\rm dec})}.
\end{eqnarray}
Following Eq.\eqref{ratio}, we can replace $\frac{\rho_{s}(T_{\rm dec})}{\rho_R^{\rm before}(T_{\rm dec})}=\frac{T_{\rm S-dom}}{T_{\rm dec}}$ and get,
\begin{eqnarray}
     \mathcal{D}&=& 1+\frac{T_{\rm S-dom}}{T_{\rm dec}}.
     \label{dilution_factor}
\end{eqnarray}

\noindent Note that, $\mathcal{D}=1$ denotes no entropy injection.

The decay of $S$ injects entropy and energy into the radiation bath, causing the temperature to rise from $T_{\rm dec}$ to a higher temperature $T_{\rm reh}^S$. For an instantaneous decay, entropy conservation gives us the relationship between the temperature change and the dilution factor. The entropy per comoving volume scales as $\mathcal{S} \propto T^3$, therefore:

\begin{equation}
    \mathcal{D} = \frac{\mathcal{S}_{\rm after}}{\mathcal{S}_{\rm before}} = \left(\frac{T_{\rm reh}^S}{T_{\rm dec}}\right)^3.
\end{equation}

Therefore, the $T_{\rm reh}^S$ is given by:

\begin{equation}
    T_{\rm reh}^S = T_{\rm dec} \times \mathcal{D}^{1/3} = T_{\rm dec} \times \left(1 + \frac{T_{\rm S-dom}}{T_{\rm dec}}\right)^{1/3}.
\end{equation}

\noindent Since $T_{\rm S-dom} \gg T_{\rm dec}$ by assumption, the dilution factor $\mathcal{D}$ is large, and consequently $T_{\rm reh}^S \gg T_{\rm dec}$. This reheating effect is crucial for understanding the thermal history after $S$ decay and impacts calculations of relic abundances for any species present during this era.

\section{Gravitational Wave Signals}
\label{GW_analysis}
Before delving into the details of GW analysis, we first identify the allowed parameter space when a DW annihilation is followed by an era of MD. We try to keep our analysis model independent and hence we choose four free parameters: (a) DW surface tension, $\sigma$, (b) bias in the scalar potential, $V_{\rm bias}$, (c) scalar decay width, $\Gamma_S$ and (d) amplitude of scalar curvature perturbation, $A_s$. We apply the constraints coming from the life time of the long-lived scalar on top of the constraints that already exist on DW allowed parameter space as show in Fig. \ref{fig:DW_constraints}. In addition, we also include the constraints on the amplitude of scalar curvature perturbation, $A_s$. We first demand that the era of MD is only induced after the DWs have vanished, $i.e~ t_{S-\rm  dec}>t_{\rm ann}$, in addition, we want $S$ domination to happen before it decay, $i.e~ t_{S-\rm dec}>t_{S-\rm dom}$. Moreover, we also want $S$ to decay before BBN  $i.e~ t_{S-\rm dec}<t_{\rm BBN}$. Finally, we want the minimum value $A_s$ to be smaller than $A_s^{\rm min}<10^{-2}$ to avoid the overproduction of PBH \cite{Harada:2016mhb,Ballesteros:2019hus,DeLuca:2021pls,Sasaki:2018dmp} and we also consider $A_s^{\rm min}>A_s^{\rm CMB}\simeq 10^{-9}$ \cite{Planck:2015fie}. 

\begin{figure*}[htbp!]
    \centering
    \includegraphics[width=0.4\linewidth]{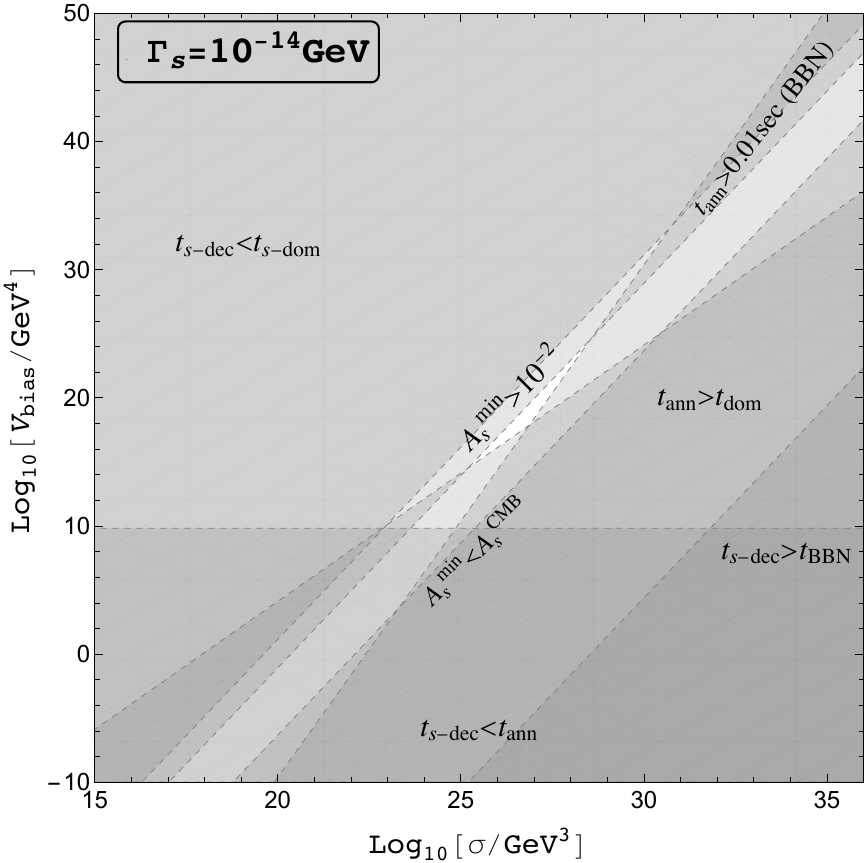}
    \includegraphics[width=0.4\linewidth]{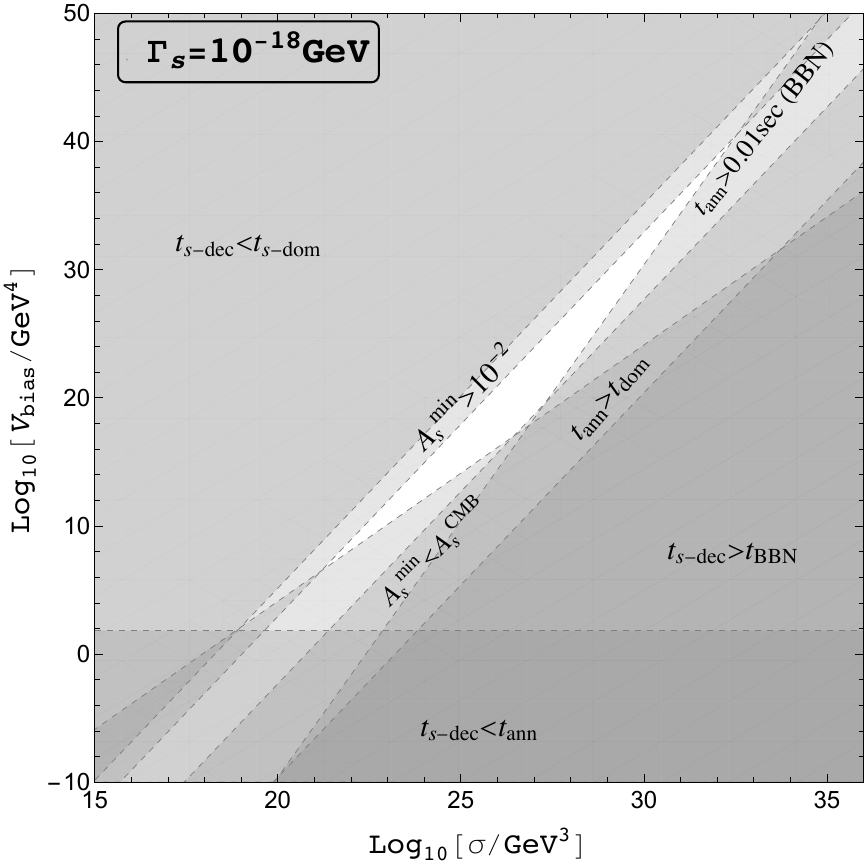}
    \includegraphics[width=0.4\linewidth]{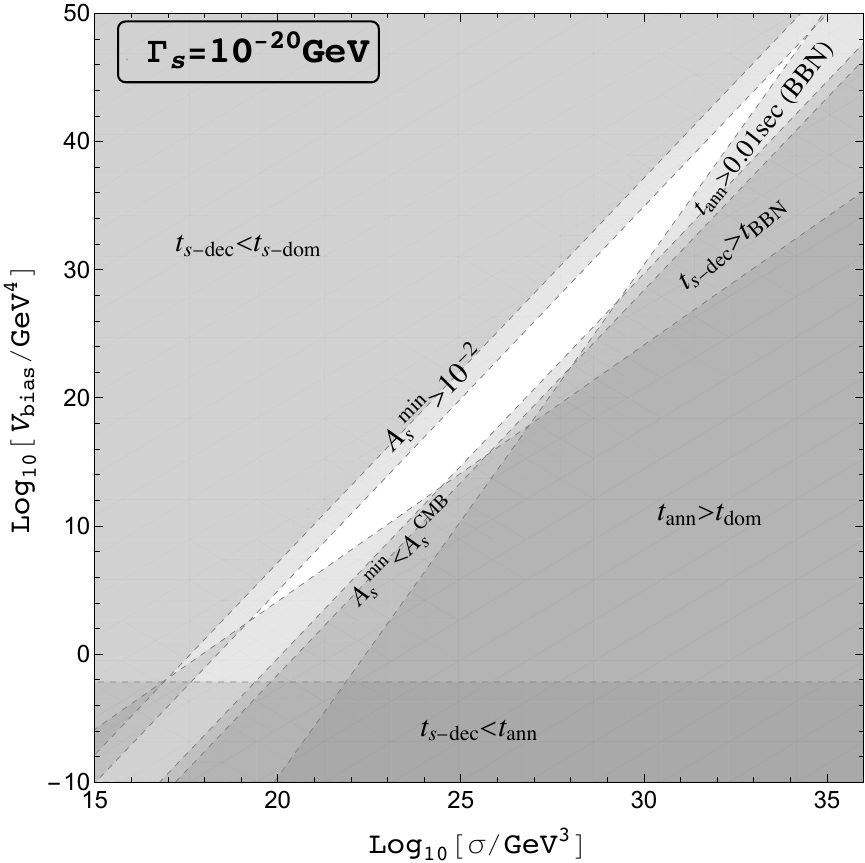}
    \includegraphics[width=0.4\linewidth]{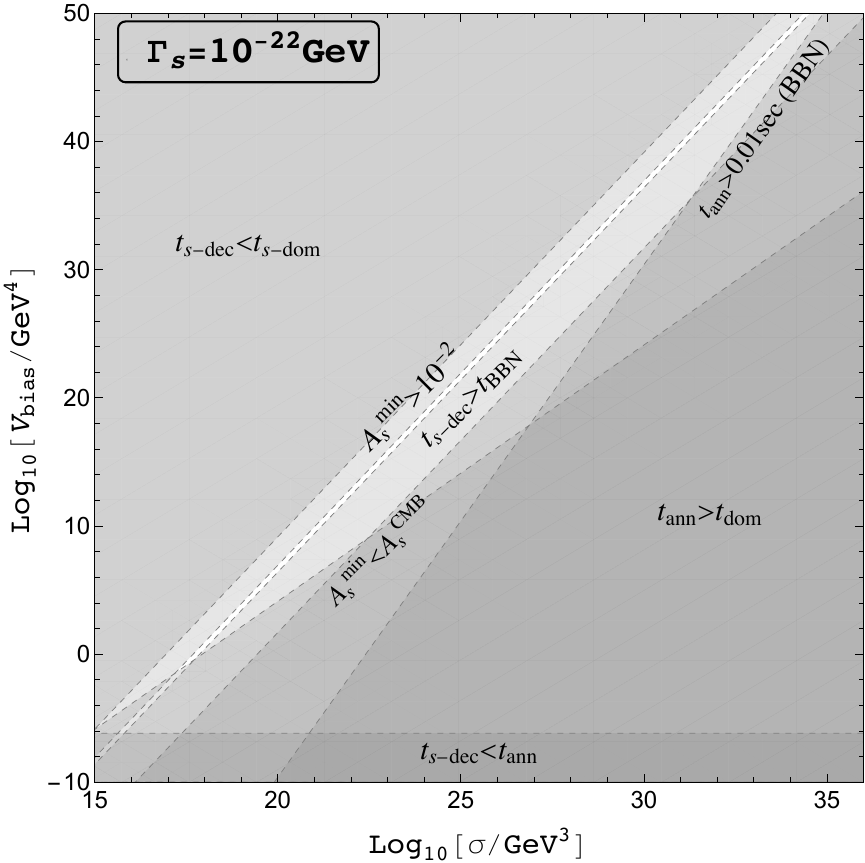}
    \caption{ Parameter space in $V_{\rm bias}-\sigma$ plane for four representative values of $\Gamma_S=10^{-14}~\rm{GeV}$, $\Gamma_S=10^{-18}~\rm{GeV}$, $\Gamma_S=10^{-20}~\rm{GeV}$, and $\Gamma_S=10^{-22}~\rm{GeV}$ respectively. Gray shaded regions show the parameter space ruled out by physical constraints.}
    \label{fig:constraints}
\end{figure*}

In addition to the constraints shown in Fig. \ref{fig:DW_constraints}, we now include these extra constraints in Fig. \ref{fig:constraints} in $V_{\rm bias}-\sigma$ plane for four representative values of $\Gamma_S$. Looking at the plots in Fig. \ref{fig:constraints}, it is clear that an induced early MD following a DW annihilation shrinks the allowed parameter drastically. In the top left panel of Fig. \ref{fig:constraints}, we fix $\Gamma_S=10^{-14}$ GeV and we observe a very narrow allowed region (shown in white), here  a significant amount of parameter space is constraint from the  overproduction of PBH, DW overabundance and DW annihilation after BBN. Since $A_s^{\rm min}$ following Eq. \eqref{eq:asbound}, depends on the duration of MD, $i.e.~ T_{\rm dec}~\rm{and}~T_{S-\rm dom}$, decreasing $\Gamma_S$ relaxes both PBH overabundance constraints as well as $A_s^{\rm min}>A_s^{\rm CMB}\simeq 10^{-9}$ resulting in a less restrictive (wider white region) parameter space. This behaviour is visible in top right and bottom left panels of Fig. \ref{fig:constraints}. Interestingly, decreasing $\Gamma_S$ further to $10^{-22}$ GeV, the BBN constraints on life time of $S$ becomes more stringent, making the parameter space shrink again. This observation restricts our choices of $\Gamma_S$ roughly in between $10^{-14}~{\rm GeV}$ to $10^{-22}~{\rm GeV}$.

\subsection{Gravitational wave from domain walls}
In this section, we examine GW production from DW annihilation and demonstrate how the resulting GW spectrum is modified by an early MD era and the subsequent transition back to radiation domination. We start with the assumption that the DW annihilate instantaneously ($t=t_{\text{ann}}$) during the radiation-dominated era, the peak frequency $f_{p}$ and peak energy density spectrum $\Omega_{p}h^2$ of GW at present can be expressed as \cite{Saikawa:2017hiv, Chen:2020wvu}
 \begin{widetext}
 \begin{align}\label{fpeak}
     f_{p}&\simeq 3.75\times10^{-9}~\text{Hz}\times C_{\rm ann}^{-1/2}\mathcal{A}^{-1/2}\bigg(\frac{10^3~\text{GeV}}{\mathcal{\sigma}^{1/3}}\bigg)^{3/2}\bigg(\frac{V_{\text{bias}}^{1/4}}{10^{-3}~\text{GeV}}\bigg)^{2}\,,\nonumber\\
     \Omega_{p}h^2&\simeq 5.3\times10^{-20}\times\tilde{\epsilon}_{\text{GW}}~C_{\rm ann}^{2}\mathcal{A}^{4}\bigg(\frac{\sigma^{1/3}}{10^3~\text{GeV}}\bigg)^{12}\bigg(\frac{10^{-3}~\text{GeV}}{V_{\text{bias}}^{1/4}}\bigg)^{8},
 \end{align}
 \end{widetext}
 where $\tilde{\epsilon}_{\text{GW}}\simeq0.7$~\cite{Hiramatsu:2013qaa} denotes the fraction of energy radiated into GW and can be regarded as a constant in the scaling regime. 


The GW spectrum from DWs characteristically exhibits a broken power-law form. The peak frequency is set by the annihilation time, while the peak amplitude is determined by the DW energy density, as shown in Eq. \eqref{fpeak}. However, if a MD phase intervenes between DW annihilation and the onset of standard radiation domination, both $\Omega_{\rm peak}$ and $f_{\rm peak}$ are modified. The peak amplitude is suppressed by the entropy dilution factor, while the peak frequency is redshifted according to the expansion history during and after the MD epoch. Incorporating these modifications, we describe the GW spectrum using the broken power-law parametrization introduced in Refs.\cite{Caprini:2019egz, NANOGrav:2023hvm}:
\begin{eqnarray}
 \Omega_{\rm GW}h^2_{} = \frac{\Omega_p h^2}{\mathcal{D}^{4/3}} \frac{(a+b)^c}{\bigg{(}b \left({\mathcal{D}^{1/3}\frac{f}{f_p}}\right)^{-a / c}+a \left(\mathcal{D}^{1/3}\frac{f}{f_p}\right)^{b / c}\bigg{)}^c} \ ,
\label{eq:spec-par}
\end{eqnarray}
where $a$, $b$ and $c$ are real and positive parameters. Here the low-frequency slope\footnote{In the presence of an early MD, the lower frequency slope should also see a slight modification as discussed in  \cite{Ghoshal:2026ros}. Since it does not effect our main result significantly, we do not include it in our analysis.} $a = 3$ can be fixed by causality, while numerical simulations suggest $b \simeq c \simeq 1$~\cite{Hiramatsu:2013qaa}.

\begin{figure}[htbp!]
    
    \includegraphics[width=0.85\linewidth]{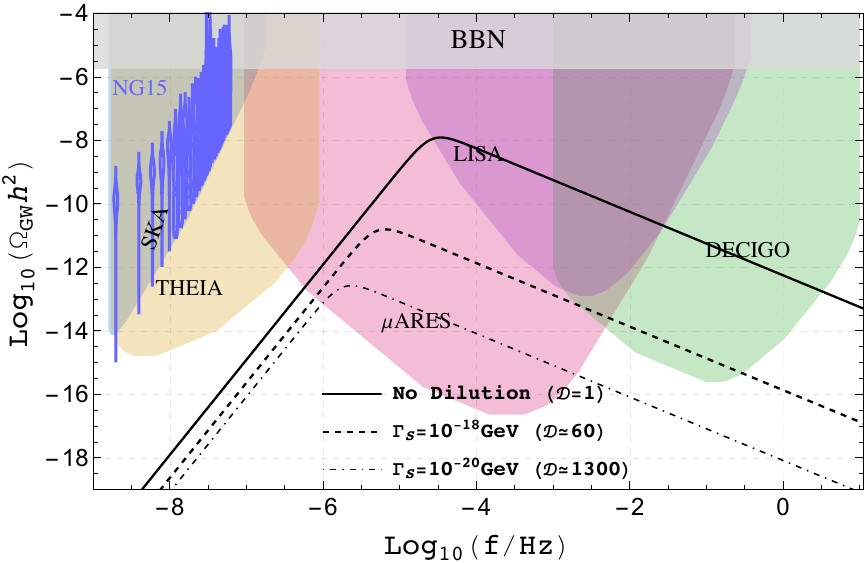}
    \caption{The GW spectrum produced from the annihilation of DWs. The effect of entropy injection is show by dashed and dot-dashed lines. The plot is made for a fixed value of $\sigma=5\times10^{22}~\rm{GeV}^3$  and $V_{\rm bias}=10^{10}~ \rm{GeV}^4$.}
    \label{fig:DW_emd}
\end{figure}

In Fig. \ref{fig:DW_emd}, we first present the power-law integrated sensitivity curves \cite{Thrane:2013oya} of the future GW detectors ET~\cite{Punturo:2010zz}, LISA~\cite{LISA:2017pwj}, DECIGO~\cite{Kawamura:2020pcg}, $\mu$Ares~\cite{Sesana:2019vho}, SKA~\cite{Janssen:2014dka}, and THEIA~\cite{Garcia-Bellido:2021zgu} which are evaluated following Eq.~\eqref{eq:spec-par} by calculating the signal-to-noise ratio (SNR)~\cite{Maggiore:1999vm,Allen:1997ad}
\begin{equation*}
    \varrho=\left[n_{\mathrm{det}} t_{\mathrm{obs}} \int_{f_{\min }}^{f_{\max }} d f\left(\frac{\Omega_{\text {signal }}(f)}{\Omega_{\text {noise }}(f)}\right)^2\right]^{1 / 2} \; ,
    \label{eq:SNR}
\end{equation*}
where $n_{\rm det} = 1$ for auto-correlated detectors and  $n_{\rm det} = 2$ for cross-correlated detectors, $t_{\rm obs}$ denotes the observational time, and $\Omega_{\text {noise }}$ represents the noise spectrum expressed in terms of the GW energy density spectrum~\cite{Schmitz:2020syl}. Integrating $(\Omega_{\rm signal}/\Omega_{\rm noise})^2$ over the sensitive frequency range of individual GW detectors, we obtain the SNRs for the GW spectra. Here we also choose
$\varrho = 10 $ as the threshold SNR. The gray shaded region (labeled BBN) represents the constraint $ \Omega_{\rm GW}^{\rm \Delta N_{\rm eff}}h^2$ \cite{Domenech:2020ssp}.

In Fig. \ref{fig:DW_emd}, we show the GW curves resulting from the DW annihilation for a fixed value of $\sigma=5\times10^{22}~\rm{GeV}^3$  and $V_{\rm bias}=10^{10}~ \rm{GeV}^4$. The choice presented
here is merely one illustrative example that remains consistent with all the constraints shown in Fig. \ref{fig:constraints}. This parameter choice serves as an illustrative example, analogous results can be obtained for any other values of $\sigma$ and $V_{\rm bias}$ within the allowed region of Fig. \ref{fig:constraints}. The GW spectrum (following Eq. \eqref{eq:spec-par}) clearly reflects the influence of early MD and the subsequent return to radiation domination. As expected, decreasing 
$\Gamma_S$ extends the scalar's lifetime, thereby increasing entropy injection into the thermal bath and amplifying the dilution factor 
$\mathcal{D}$ (see Eq. \eqref{dilution_factor}). This entropy release suppresses the peak amplitude of the GW signal while simultaneously red-shifting the peak frequency in accordance with the modified expansion history during and after the MD phase. This is seen from the black dashed and black dot-dashed line of Fig. \ref{fig:DW_emd}. For reference, we also include the undiluted GW spectrum (solid black line) corresponding to $\mathcal{D}=1$.

\subsection{ Enhanced induced gravitational waves}
\label{EIGW}

It is well known that during the early Universe's evolution, primordial scalar curvature perturbations can indirectly generate GW through second-order effects in general relativity. While at first order, they cannot directly produce the transverse-traceless tensor modes that constitute GW, extending the perturbation theory to second order reveals that products of these first-order scalar modes can source tensor perturbations \cite{Matarrese:1992rp,Matarrese:1993zf,Matarrese:1997ay,Mollerach:2003nq,Ananda:2006af,Baumann:2007zm,Domenech:2021ztg,Roshan:2024qnv}. As modes enter the horizon during radiation domination, scalar perturbations interact non-linearly to produce GW through their coupling. The amplitude of these induced GWs depends quadratically on the scalar power spectrum $\Omega_{\rm GW} \propto A_s^2$ \cite{Domenech:2021ztg}, with $A_s$ denoting the amplitude of the scalar curvature perturbations . The resulting IGW spectrum carries information about the scalar perturbation spectrum at small scales and provides an independent window into early Universe physics.

Moreover, presence of an early MD era can intensify the production of these IGWs \cite{Assadullahi:2009nf,Baumann:2007zm,Jedamzik:2010dq,Alabidi:2013lya,Kohri:2018awv,Inomata:2019ivs,Inomata:2019zqy,Papanikolaou:2020qtd,Domenech:2020ssp,Dalianis:2020gup,Domenech:2021ztg,Das:2021wad,Eggemeier:2022gyo,Fernandez:2023ddy,Chianese:2025mll} and the degree of enhancement depends on whether perturbations evolve linearly or non-linearly. For the non-linear case, density perturbations grow and can lead to the formation of non-linear structures \cite{Fernandez:2023ddy} (here $S-$halos).  The scale of non-linearities, $k_{\rm NL}(T)$ at the time of $S$ decay is estimated by \cite{Fernandez:2023ddy}
\begin{eqnarray}
  k_{\rm NL}(T_{\rm dec})\sim 1.7 A_s^{-1/4} \mathcal{H}(T_{\rm dec}).
  \label{kNL}
\end{eqnarray}

\noindent In addition, the transition from MD era to the subsequent radiation-dominated era further complicates predictions, as the transition dynamics affect the final spectrum \cite{Fernandez:2023ddy}. 

In a linear regime, i.e. for $k<k_{\rm NL}$, the IGW energy density still scales as $\Omega_{\rm GW} \propto A_s^2$ but with a relatively larger coefficient in front of $A_s$ \cite{Inomata:2019zqy} in comparison to the standard radiation dominated era. The simulation done in \cite{Fernandez:2023ddy}, does not shows a sufficient enhancement of the GW spectrum over the perturbative prediction in a linear regime.

In the non-linear regime, i.e. for $k>k_{\rm NL}$, as shown in Ref.~\cite{Fernandez:2023ddy}, the IGW generation is dominated by the largest structures that form at the end of the MD era.  Hence a longer MD is preferred in the non-linear regime. Specifically, the largest non-linear scale must enter the Hubble horizon during the early MD phase,  $i.e~k_{\rm NL}(T_{\rm dec})<{\cal H}(T_{S-\rm dom})$. This inequality translates into a lower bound on $A_{\rm s}$ given by \cite{Fernandez:2023ddy, Chianese:2025mll}
\begin{equation}
    A_{\rm s}^{\rm min}\equiv \alpha^4 \left(\frac{a_{\mathrm{dom}}}{a_{\mathrm{dec}}}\right)^2 \approx 9 \left(\frac{T_{\mathrm{dec}}}{T_{\mathrm{dom}}}\right)^2\,.  \label{eq:asbound}
\end{equation}


We adopt the state-of-the-art fit from Ref.~\cite{Fernandez:2023ddy} for the IGW spectrum induced by a scale-invariant power spectrum at the end of the early MD epoch. Expressing the GW spectrum today in terms of frequency we get \cite{Chianese:2025mll}, 
\begin{equation}
    \Omega_{\rm GW} h^2 = 4.2\times 10^{-5}\, A_{s}^{11/8} \left( \frac{f}{f_\mathrm{high}} \right)^{3/2} \,, \label{eq:Gw_present} 
\end{equation}
where we account for the redshifting factor assuming the SM effective degrees of freedom,
\begin{equation}  
    f_{\rm high} \simeq 6.4 \times 10^{-5} \text{ Hz}\left(\frac{A_{\rm s}}{10^{-5}}\right)^{-1/4} \left(\frac{T_{\mathrm{dec}}}{10\, \text{GeV}}\right) \,,
    \label{eq:fhi}
\end{equation}  
\begin{equation}  
    f_{\rm low}\simeq 3.8 \times 10^{-6} \text{ Hz}\left(\frac{T_{\mathrm{dec}}}{10\, \text{GeV}}\right)\,. 
    \label{eq:flow} 
\end{equation}
Here, $f_{\rm low}$ and $f_{\rm high}$  cut-offs corresponds to $k_{\rm low}\approx 15 {\cal H}_{\rm dec}$ and $k_{\rm high}\approx 9 k_{\rm NL}(\tau_{\rm dec})\approx 14 \mathcal{H}_\mathrm{dec} / A_\mathrm{s}^{1/4}$, respectively. The lower limit arises from numerical resolution constraints~\cite{Fernandez:2023ddy}, while the upper limit stems from potential artifacts in non-relativistic $N$-body simulations related to coherent effects, corresponding to the light-crossing time of the largest halos.
However, a smooth spectral decay is generally expected rather than an abrupt cutoff at $k_{\rm high}$. Studies~\cite{Dalianis:2020gup, Eggemeier:2022gyo, Fernandez:2023ddy, Dalianis:2024kjr} indicate a high-frequency power-law decay  $\Omega_\mathrm{GW}\propto 1/f^n$ with $n \gtrsim 1$, though fully relativistic simulations are necessary to confirm this \cite{Adamek:2016zes}. We adopt  $\Omega_\mathrm{GW}\propto f^{-1}$ for $f>f_{\rm high}$for simplicity. At sufficiently low frequencies, linear perturbation theory predictions are recovered.

\begin{figure}[htbp!]
    
    \includegraphics[width=0.85\linewidth]{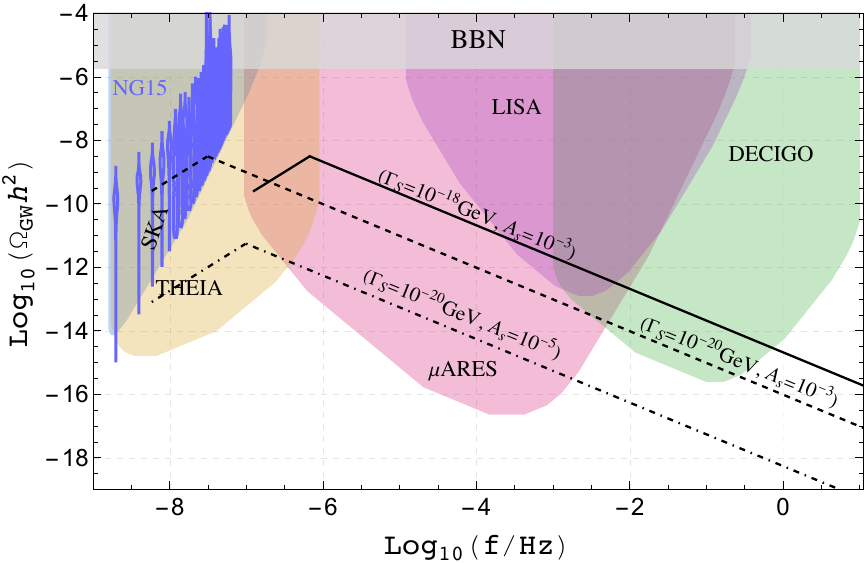}
    \caption{The enhanced primordial induced GW spectrum resulting from the early matter domination driven by $S$. The plot is made for different combination of $\Gamma_S$ and $A_s$.}
    \label{fig:IGW}
\end{figure}

 Following Eqs. \eqref{eq:Gw_present}, \eqref{eq:fhi}, and \eqref{eq:flow}, it is clear that the two parameters that controls the IGW amplitude and frequency are $\Gamma_S$ and $A_s$, hence
in Fig. \ref{fig:IGW}, we show the variation of $\Omega_\mathrm{GW}h^2$ with frequency $f$ for different values of $\Gamma_S$ and $A_s$. Comparing the solid and the dashed line one finds that for a fixed value of $A_s=10^{-3}$, changing the decay width $\Gamma_S$ from $10^{-18}$ to $10^{-20}$, shift the spectrum towards the lower frequency while keeping the $\Omega_\mathrm{GW}h^2(f_{\rm high})$ same, this is because at  $f_{\rm high}$, the $\Omega_\mathrm{GW}h^2$ remains independent of $\Gamma_S$. A significant shift in $\Omega_\mathrm{GW}h^2(f_{\rm high})$ is observed when $A_s$ is changed from $10^{-3}$ to $10^{-5}$ for a fixed value of $\Gamma_S$ (dashed and dot-dashed lines). This behaviour is clear from Eq. \eqref{eq:Gw_present} where a strong dependence on $A_s$ is observed. These benchmarks are also consistent from all the constraints show in Fig \ref{fig:constraints}.

\section{Summary}
\label{sec:summary}
\begin{figure*}
    \includegraphics[width=0.85\linewidth]{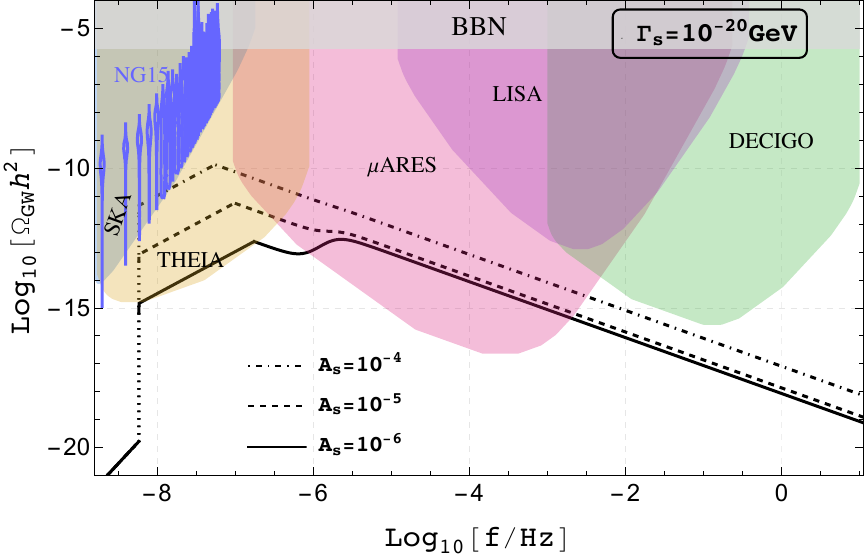}
    \caption{ The combined GW spectrum arising from the DW annihilation and enhanced IGW for a fixed value $\sigma=5\times10^{22}~\rm{GeV}^3$  and $V_{\rm bias}=10^{10}~ \rm{GeV}^4$ respectively.}
    \label{summary}
\end{figure*}

Fig. \ref{summary} summarizes our results for a scenario in which DW annihilation precedes an early MD era driven by the same scalar field whose $\mathbb{Z}_2$ symmetry breaking produced the DW initially. The predicted GW spectrum shows a double-peaked structure reflecting two distinct physical mechanisms: DW annihilation generates the higher-frequency peak, while IGW produce the lower-frequency feature. The separation between these peaks comes from the fact that the MD era enhancing the IGW signal begins only after DWs have annihilated. As evident from the spectrum in Fig. \ref{summary}, this scenario offers a striking observational signature: a single symmetry-breaking process yields GW accessible to both high-precision astrometry (THEIA) and space-based interferometry ($\mu$ARES). Detection in one frequency band would immediately indicate the presence of a complementary signal in the other. Notably, moderately enhanced scalar curvature perturbations can amplify the IGW component to the point where it overshadows the DW signal entirely.

\section{Conclusion}
\label{conclusion}

We study a minimal SM extension containing a real scalar singlet with non-trivial $\mathbb{Z}_2$ charge. While spontaneous $\mathbb{Z}_2$ breaking is known to create DWs that generate GW through annihilation, we demonstrate that the full phenomenology is significantly richer when the scalar has a long lifetime. Such a long-lived scalar necessarily induces an early MD phase following DW annihilation, which enhances IGWs from primordial perturbations. The scalar's eventual decay  dilute the DW signal via entropy injection while preserving the induced component's amplification. This produces a two distinct peaks in the GW spectrum with a remarkable property: both peaks are potentially detectable in high-precision astrometry like THEIA and space-based interferometry like $\mu$ARES. Detection by one experiment predicts observable signals for the other, establishing a powerful consistency check. Additionally, we find that moderately enhanced primordial curvature perturbations can boost the induced component to levels that completely dominate over DW's contributions.

\section{Acknowledgment}			
RR acknowledges financial support from the STFC Consolidated Grant ST/T001011/1. 
\bibliographystyle{apsrev4-1}
\bibliography{refs}

\onecolumngrid

\end{document}